\documentclass[aip,amsmath,amssymb,
preprint,
]{revtex4-1}
\usepackage{graphicx}
\usepackage{dcolumn}
\usepackage{bm}
\usepackage[utf8]{inputenc}
\usepackage[T1]{fontenc}
\usepackage{mathptmx}
\begin{document}
\setlength\abovedisplayshortskip{1pt}
\setlength\belowdisplayshortskip{1pt}
\setlength\abovedisplayskip{1pt}
\setlength\belowdisplayskip{1pt}
\title{On the spontaneous magnetization of two-dimensional ferromagnets.}

\author{D. Pescia}
\affiliation{Laboratory for Solid State Physics, ETH Zurich, 8093 Zurich, Switzerland}
\email{pescia@solid.phys.ethz.ch}%
\author{A. Vindigni}
\affiliation{Laboratory for Solid State Physics, ETH Zurich, 8093 Zurich, Switzerland.}
\affiliation{Department of Evolutionary Biology and Environmental Studies, University of Zurich, 8057 Zurich, Switzerland.}

\begin{abstract}
Ferromagnetism is typically discussed in terms of the exchange 
interaction and magnetic anisotropies. Yet real samples are inevitably affected 
by the magnetostatic dipole-dipole interaction.  
Because of this interaction, a theorem  
(R.B. Griffiths, Free Energy of interacting magnetic dipoles, {\it Phys. Rev.} \textbf{176}, 655 (1968)) 
forbids a spontaneous magnetization in, nota bene, three-dimensional bodies. 
Here we discuss perpendicularly and in-plane magnetized ferromagnetic bodies in 
the shape of a slab of finite thickness. In perpendicularly magnetized slabs, magnetic domains are energetically favored when the lateral size is sufficiently large, i.e. there is no spontaneous magnetization. For in-plane magnetization, instead, spontaneous magnetization is possible below a critical thickness which, in real thin films, could be as small as few monolayers. At this critical thickness, we predict a genuine phase transition to a multi-domain state. These results have implications for two-dimensional ferromagnetism.   
\end{abstract}
\date{\today}
\maketitle
A situation familiar to  ferromagnetism foresees that, below the Curie temperature, the graph of the the free-energy as a function of the 
magnetization has a flat portion\cite{Ising} (graph ''a'' in Fig.1). This flatness defines a situation in which the magnetization can acquire a value between zero and a so called ''spontaneous magnetization'' $\pm {M}_0$ without any change in the free energy. This ''flatness'' is a property of the thermodynamic limit, i.e. of infinite bodies. In finite bodies, the free energy assumes a shape that resembles the graph ''b'' in Fig.1, $\pm M_0$ being the values at which the free energy has minima. Systems with spontaneous magnetization are, for example, the Ising and classical 
Heisenberg ferromagnets in three dimensions (3D)\cite{Ising,Jurg}, the 2D-Ising 
model or the 2D-planar and classical 2D-Heisenberg models with symmetry breaking single ion interactions\cite{Jose}. In real ferromagnetic 
bodies, however, the inevitable dipole-dipole interaction, originating within 
Maxwell equation of magnetostatics, must be considered alongside the main exchange interaction (of purely quantum mechanical origin) and the single ion magnetic anisotropies (produced by the spin-orbit interaction). The dipole-dipole 
interaction is, typically, much weaker than the exchange interaction (by 
about two orders of magnitude). Yet, it is long-ranged, as it decays only 
with the third power of the distance between two magnetic moments. Because of 
the dipole-dipole interaction, a theorem, proved by Griffiths\cite
{Griff} for bodies with linear dimension $L$ approaching infinity along all 
three spatial dimensions, implies that any non-zero magnetization produces an 
increase of the free energy, i.e. the graph of the free energy as a function of $M$ has a minimum at $M=0$ at any temperature (graph ''c'' in Fig.1). Accordingly, ferromagnetic order can only be local and, globally, the spontaneous magnetization is exactly vanishing.\\
This no-spontaneous magnetization rule is somewhat similar to the absence of 
long-range order foreseen for the isotropic 2D-planar and 2D-Heisenberg ferromagnets\cite
{Mermin} but it refers, remarkably, to a 3D-body. In fact, it appears that an 
important assumption underlying Griffith's theorem is the size of the body 
approaching infinity along all three spatial dimensions. In this paper, we 
discuss ferromagnetism in the presence of exchange, magnetic anisotropy and 
dipole-dipole interaction but in a slab-geometry, where only two spatial 
dimensions are allowed to increase and the third is assigned a finite thickness. 
Our results should be relevant for discussing ferromagnetism in the new class of 
monolayer thin materials obtained by mechanical exfoliation\cite{Gong,Huang,Burch,Li,Novo,Dai}. They are known to be 
perfectly flat over large distances and have been shown to be vertically 
engineerable\cite{Wua}. As experiments are often analyzed in term of abstract 
models, exact theorems such as Griffiths's one\cite{Griff} or scaling 
arguments such as those presented here should allow experiments to distinguish those features that are general and universal from those ones that originates from less known details of a sample (such as defects).\\
\textbf{Perpendicular Magnetization.} We first analyze the situation of perpendicular magnetization. In the state of  
spontaneous perpendicular magnetization, all magnetic moments in the slab point 
along one of the two $z$-directions perpendicular to the slab (Fig.2a), e.g. the $+z$-direction. In Fig.2a, this state is rendered with a white color and the magnetization vector with absolute value $M_0$ is given by a black arrow. $M_0$ represents the magnetic moment per volume of the unit cell, i.e. $M_0\doteq \frac{g\cdot \mu_B\cdot S}{a^3}$ (with $S$ being the spin in units of $\hbar$ and $a$ the lattice constant). A possible, elementary state of vanishing spontaneous magnetization is shown in Fig.2b. One half of the slab is still filled with magnetic moments pointing upwards ''$\uparrow$'' but the other half (gray in Fig.2b) contains magnetic moments pointing downwards ''$\downarrow$'' (indicated as state $\uparrow\downarrow$ henceforth). Assuming that Griffith's theorem is valid for the perpendicular magnetization in the slab geometry as well, the  $\uparrow\downarrow$ state should have a lower total energy than the state of spontaneous uniform magnetization (indicated as state $\uparrow\uparrow$ henceforth).  We, therefore, proceed to compare the energy of the two states in a situation where the size $L$ of the slab is much larger than its thickness $d$.\\
\textit{Magnetostatic energy $E_M$.} The magnetostatic energy $E_M$ for the perpendicular magnetization configuration is most appropriately computed as the  energy of the interacting Amp\`erian effective currect densities $\vec 
\nabla\times \vec M(\vec r)$ resulting from  the magnetization vector $\vec 
M(\vec r)$ (see Section I of the Supplementary Material (SM)). The current 
density vectors arising in the $\uparrow\uparrow$ and $\uparrow\downarrow$ configurations are summarized by red arrows in Fig.2. We find that the formation of the two 
neighboring domains with opposite magnetization lowers the total magnetostatic 
energy and is thus the driving force behind the suppression of the spontaneous 
magnetization, i.e. $E_M(\uparrow\downarrow)-E_M(\uparrow\uparrow)$ is negative. The self-energies of the Amp\`erian currents circulating along the perimeter of the slab cancel out exactly from 
$E_M(\uparrow\downarrow)-E_M(\uparrow\uparrow)$. Their mutual interaction provides terms of the order $L\cdot d^2$. The leading contribution to $E_M(\uparrow\downarrow)-E_M(\uparrow\uparrow)$ is the self-energy of the current flowing along the wall that separates the domains. Assuming that the domain wall  has a finite thickness $w$, the leading contribution writes (Section I, SM), in the limit  $d\!<\!<w\!<\!<\!L$ 
\begin{equation}
E_M(\uparrow\downarrow)\!-\!E_M(\uparrow\uparrow)\!\approx \!\frac{L\!\cdot\!d}{a^2}\!\cdot\!\left[-\frac{2}{\pi}\cdot (\Omega\!\cdot\!\frac{d}{a})\!\cdot\!  \ln\frac{L}{w}\right]+ \!{\cal O}(\frac{d^2\!\cdot\! L}{a^3})
\end{equation} 
In Eq.1, $L\!\cdot \!d$ is the surface of the domain wall. $\Omega\doteq \frac{\mu_0}{2}\cdot M_0^2\cdot a^3$ is a parameter used for 
expressing the strength of the magnetostatic energy per unit cell. As an example, $\Omega$ for metallic Fe amounts to $\approx 0.28$ meV (see Section I of SM)\cite{omega}. The slab model (see Section I of SM) shows explicitly that the relevant coupling constant entering  $E_M(\uparrow\downarrow)\!-\!E_M(\uparrow\uparrow)$ is not $\Omega$ itself but $\Omega\cdot\!\frac {d}{a}$, i.e. the characteristic magnetostatic energy per unit surface cell. The logarithmic term in Eq. 1 provides, formally, a divergence of $E_M(\uparrow\downarrow)\!-\!E_M(\uparrow\uparrow)$ with the size $L$. It is universal in the sense that it does not depend on the exact geometry of the wall separating the domains: both the shape of this wall and the exact shape of slab contribute only terms of the order $\Omega\!\cdot\!L\cdot\! d^2$.\\ 
\textit{Energy of the wall between $\uparrow$-and $\downarrow$-domains.} The formation 
of a domain wall in the $\uparrow\downarrow$-state increases the total energy of the ferromagnetic slab and therefore promotes the state of spontaneous 
magnetization. Within the wall, the magnetic moments rotate away from the 
$z$-direction. For simplicity, we assume the wall to run parallel to the $y$-direction and the rotation to take place along the $x$-direction. Let the 
rotation be characterized by an angle $\theta$, which increases from $0$ to $\pi$
when moving along $x$-within the wall. The misalignement is associated, in the 
first place, with an increase of the single ion magnetic anisotropy energy that 
favors the perpendicular magnetization, introduced conceptually by N\'eel\cite
{Neel} and computed for the first time from first principles by Gay and Richter
\cite{Gay} for the monolayer of Fe. This term originates from the 
breaking of translational symmetry perpendicular to the slab plane. In ultrathin 
slabs, it is only weakly dependent on $d$\cite{GR} as it arises from the two 
surfaces bounding the slab. Using the convention of Ref.\cite{Gay}, we write this term as $-\lambda\cdot \cos^2\theta(x)$, the parameter $\lambda$ to be intended, as in Ref.\cite{Gay}, as an energy per surface unit cell. 
For $\lambda$, a value of $\approx 0.4$ meV is reported\cite{Gay,Shi} for the one monolayer of Fe.\\
The N\'eel-anisotropy is not the only contribution to the magnetic anisotropy affecting the magnetic moment rotation in the wall. The magnetostatic energy itself favors the magnetic moment to lie within the slab plane and contributes a term  $+\Omega\!\cdot\!\frac{d}{a}\cdot\! \cos^2\theta(x)$ (Section II of SM). The coefficient of this contribution scales with $d$, see Section II of SM and Ref. \cite{GR}, in contrast to $\lambda$.\\
Finally, for the building of the wall, one must also consider that the 
misalignment of two neighboring magnetic moments at the sites $x$ and $x\pm a$ 
increases the energy by\cite{Small} $J\cdot S^2\cdot \cos\left(\theta(x\pm 
a)-\theta(x)\right)$, $J$ being the exchange coupling energy per spin couple. 
For bulk Fe, Ref.\cite{Small} estimates $J\approx 48$ meV. The total energy of a 
wall is proportional to its surface $L\cdot d$ multiplied by the geometric mean 
of the relevant coupling constants (see Section III of SM for a short reminding 
on how the term $+\Omega\!\cdot\!d\!\cdot\! \cos^2\theta(x)$ is embedded into the wall energy): 
\begin{equation} \propto \frac{L\cdot d}{a^2}\!\cdot\!\left[2\cdot\!\sqrt{\left(\lambda-\Omega\!\cdot\!\frac{d}{a}\right)\!\cdot\!2\!\cdot\! J\cdot S^2}\right]
\end{equation}
\textit{Absence of spontaneous magnetization and crossover length.} Comparing the domain wall energy cost to the 
magnetostatic energy gain, we recognize that the logarithmic term always favors 
the building of domains for sufficiently large $L$. Accordingly, Griffith's theorem about the absence of spontaneous magnetization in the thermodynamic limit holds true in the slab geometry with perpendicular magnetization.\\
One interesting outcome of our argument is the estimate of the cross-over length $L_c$ at which a slab will transit from a  monodomain state to a multi-domain 
state, resulting from equating the magnetostatic energy gain to the wall energy:  
\begin{equation}
L_c\approx w\cdot e^{\frac{\pi\cdot \sqrt{\left(\lambda-\Omega\!\cdot\! \frac{d}{a}\right)\cdot 2\cdot J\cdot S^2}}{\Omega\!\cdot\!\frac{d}{a}}}
\end{equation}
There are three aspects of this result that deserve amplification. First, if we 
insert the values for $J$, $\Omega$ and $\lambda$ given previously, one 
recognizes that $L_c$ assumes astronomically large values for the monolayer 
limit (In Section III of SM we find that the ''monolayer limit'' corresponds, in the slab model, to  $d\!\approx\!a$; in this limit, 
the expression  $L_c$ is consistent with known results\cite{Pighin}).\\
Second, one recognizes a  threshold thickness $\frac{d_R}{a}= \frac{\lambda}{\Omega}$ at which, formally, the argument of the exponential function vanishes. $d_R$ is in the subnanometer range\cite{Bader,Shi}. We therefore propose that $L_c$ decreases exponentially with $d$ and, toward $d_R$, it assumes values of few tens of micrometers. These are the lateral lenghts over which exfoliated two-dimensional magnets are believed to be almost perfectly flat\cite
{Gong,Huang,Burch,Li,Novo,Dai}. Accordingly, a sequence of exfoliated 
samples with suitable thickness and with increasing lateral size $L$ should allow an insight into the yet unexplored  mechanism of penetration of magnetic domains in two-dimensional ferromagnetic elements as a function of their size 
$L$. Some preliminary results in this direction were reported in Ref.\cite{Oliver} on epitaxially grown ultrathin films.\\
Third, the spin wave excitations produce a renormalization of the various coupling constants $J$, $\Omega$ and $\lambda$ as a function of temperature
\cite{Politi}, so that $d_R$ is itself a function of the temperature. One finds 
that $d_R(T)$ defines a line of phase transitions at which the perpendicular 
magnetization turns into the plane of the slab\cite{Bader,Shi,Politi}. 
Accordingly, the exponential decay of $L_c$ should become observable when the 
temperature $T$ is increased and $d_R(T)$ approaches the actual thickness $d$ of the slab.\\
A final comment is dedicated to the stability against a perpendicular magnetic 
field of the domain phase that should appear at sufficiently large $L$. In 
Section IV of the SM, we analyze this problem by considering the energetics of 
one stripe of reversed magnetization $-M_0$, embedded into a background with magnetization $+M_0$, subject to a perpendicularly applied field 
with strength $+B_0$. We find that the state of uniform magnetization becomes the 
energetically favored one when $B_0$ exceeds a threshold strength $B_t\propto 
\left(\mu_0\cdot M_0\cdot \frac{d}{L_c}\right)$. Far away from the 
$d_R(T)$-transition line, the threshold field 
might be as small as few $nT$. Close to the transition line we expect this field 
to be in the $mT$-range\cite{Niculin,NiculinPRB}.\\ 
\textbf{In-plane magnetization.} We now analyze the slab geometry with in-plane 
magnetization. The two configurations considered are one of uniform 
magnetization along e.g. the $+x$-direction (Fig.3a), and one where one half of the slab has magnetization along the $+x$-direction and the other half has magnetization along $-x$ (Fig.3b). In this situation, the magnetostatic energy is 
most appropriately computed as the Coulomb interaction between effective 
''charges'' produced by $\vec\nabla\cdot \vec M(\vec r)$. The charges resulting 
from the two spin configurations are indicated in Fig.3 in red. The change of 
magnetostatic energy produced by the building of in-plane domains is negative, 
i.e. the magnetostatic energy  favor a state of vanishing spontaneous 
magnetization. However, the logarithmic term 
produced by the self-energies cancel out exactly when the energies of the two 
states are subtracted, provided the wall is parallel to the 
magnetization, i.e. the wall is not charged. The remaining 
contributions provide terms proportional to $-(\Omega\cdot d)\cdot L\cdot d$ (Section V, SM). Again, there is a wall between the two domains, in which spins rotate away from the $x$-direction. We assume, for simplicity, an in-plane uniaxial anisotropy with the strength $\Lambda$ such as the one encountered in ultrathin Fe films on W(110)\cite{Grad}. This uniaxial anisotropy provides and energy barrier against rotations away from the $x$ in-plane direction. A typical value for $\Lambda$, desumed from Ref.\cite{Grad}, is $0.04$ meV per unit surface cell\cite{Grad2}. Notice that $\Lambda$ originates from the rectangular nature of the surface unit 
cell and is about two-orders of magnitude smaller than the N\'eel magnetic 
anisotropy constant forthcoming in the perpendicular magnetization configuration. 
$\Lambda$ is rather of the same order of magnitude as the quartic in-plane magnetic anisotropy constant computed e.g. in Ref.\cite{Gay}. Given $\Lambda$, the energy of the wall per wall surface unit cell is then proportional to the geometric average of the exchange coupling $J$ and $\Lambda$, by virtue of the same arguments exposed in Section 3 of SM, i.e. $\approx L\cdot d\cdot 2\!\cdot 
\sqrt{\Lambda\!\cdot\!2\cdot J\cdot S^2}$.\\ 
Equating the total energy change due to the formation of a domain wall to zero provides an estimate of the critical thickness $d_c$ below which a state of spontaneous magnetization is favored:  
\begin{equation}
d_c\propto  a\cdot \frac{\sqrt{\Lambda\cdot J}}{\Omega}
\end{equation}
As $L$ cancels out, we argue that it should be possible to find a rigorous proof of spontaneous magnetization even in the thermodynamic limit 
$L\rightarrow \infty$ in a truly 2D system with in-plane magnetization. As both states above and below $d_c$ are stable for $L\rightarrow \infty$, the transition at $d_c$ from a single-domain state 
with spontaneous magnetization to a multi-domain stripe state should be a genuine 
phase transition. Using the values for $\Lambda,J,\Omega$ introduced in this 
Letter, we obtain $d_c\approx$ five lattice constants or less. This small number means that a sample must be fabricated with uniform thickness over large lateral 
distances, in order for this transition to be observed. The new class of two 
dimensional magnets\cite{Gong,Huang,Burch,Li,Novo,Dai} might provide this kind of precision. A final remark: provided that $\Omega$, 
$J$, and $\Lambda$ renormalize slightly differently with temperature, we might expect a $d_c(T)$ line of phase transitions which can also be crossed at a fixed thickness $d$ by varying the temperature. This situation would represent the analogon to a similar phase transition observed in perpendicularly magnetized films\cite{Niculin}.\\\\ 
\textbf{Supplementary Material.} Details of the computations used to obtain Eq.1,2,3 and 4 are given in the ''Supplemental Material''.\\
\textbf{Author contribution:} D.P. wrote the paper and A. V. contributed to the scientific content and to the writing.\\   
\textbf{Competing Interests:} The authors declare that they have no
competing financial interests.\\
\textbf{Data availability statement:} The informations that support the findings of this study are available within the article and its supplementary material.

\clearpage

\begin{figure}
\includegraphics*[width=0.5\textwidth]{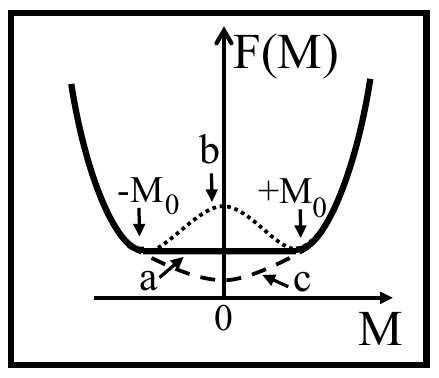}
\caption{The free energy as a function of $M$. a: the graph for an infinite ferromagnetic body with spontaneous magnetization has a flat portion between $\pm M_0$ below the Curie temperature, see Ref.\cite{Ising}. b: for a finite ferromagnetic body, the free energy has minima at $\pm M_0$. c: A theorem by Griffiths \cite{Griff} implies that the free energy has a minimum at $M=0$ at any temperature.
}
\end{figure}

\begin{figure}
\includegraphics*[width=0.5\textwidth]{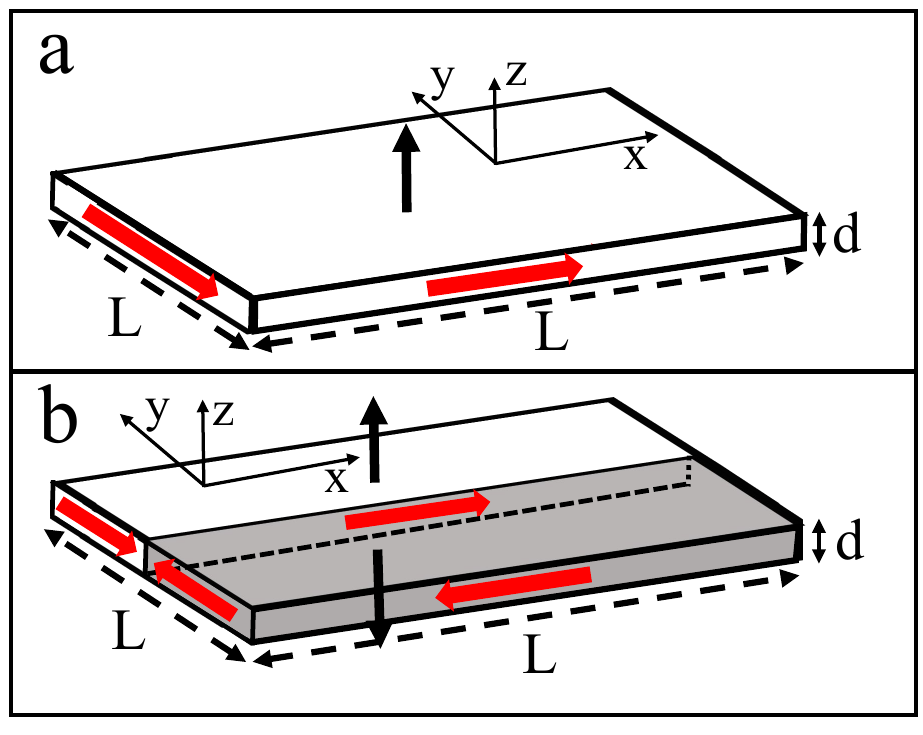}
\caption{a: the state of uniform perpendicular magnetization (represented in white) in a slab. The magnetization vector is represented by the vertical black arrow. Red arrows represent the effective current density vectors flowing along the perimeter of the slab. b: The slab is filled by two domains with magnetization vector parallel (white domain) and antiparallel (gray domain) to the vertical $z$-axis. Red arrows represent the effective current density vectors. 
}
\end{figure}

\begin{figure}
\begin{center}
\includegraphics*[width=0.5\textwidth]{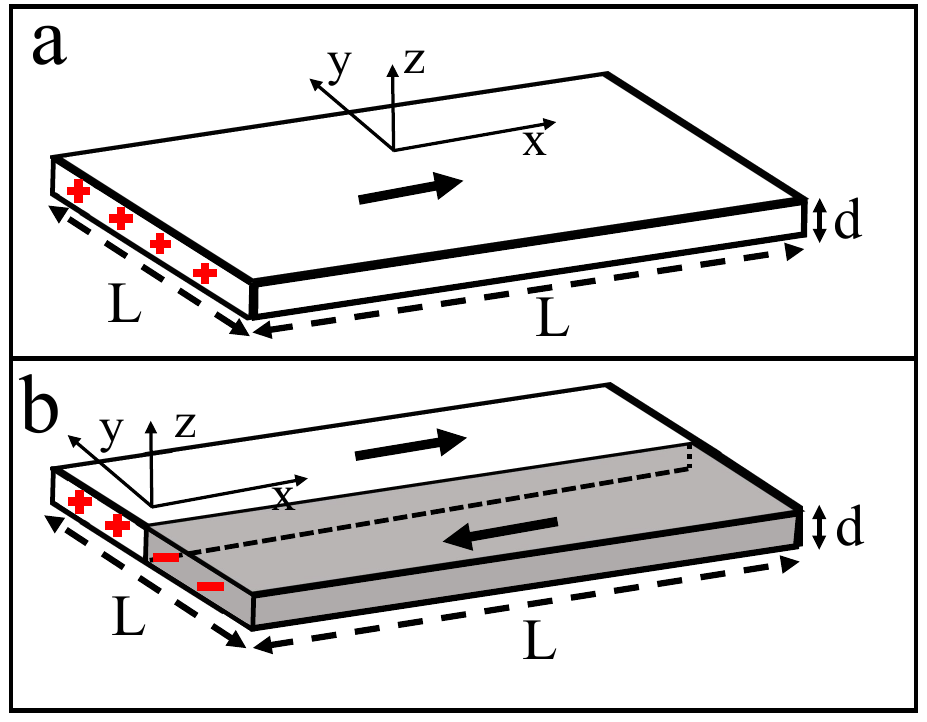}
\caption{a: the state of uniform in-plane magnetization (represented in white) in a slab. The magnetization vector is represented by the horizontal black arrow. Effective charge densities (their sign being given in red) appear along the perimeter. b: The slab is filled by two domains with magnetization vector parallel (white domain) and antiparallel (gray domain) to the horizontal $x$-axis. The sign of the effective charge densities is given in red.  
}
\end{center}
\end{figure}

\begin{thebibliography}{1}
\bibitem{omega}
In bulk Fe ($a\approx 0.28$nm) there are two magnetic moments per unit cell, each carrying a magnetic moment of about $2.2$ $\mu_B$ ($g\approx 2$, $S\approx 1.1$\cite{Small}), leading to $\Omega\approx 0.28$ meV (see also Section I of SM). For the monolayer of Fe, Gay and Richter\cite{Gay} give $\Omega\approx 0.3$ meV.
\bibitem{Grad2}
In Ref.\cite{Grad}, the anisotropy parameter is given as $0.105 \cdot \frac{\text{erg}}{\text{cm}^2}$. To convert this energy in meV per unit cell we use $1 \text{erg} = 6.242\cdot 10^{11} eV$. In the [110]-plane of bcc Fe we have 
$\frac{\sqrt{2}}{2.83^2}\cdot  \frac{\text{Atom}}{\text{Angstrom}^2}$ 
so that the magnitude of the two-fold in-plane magnetic anisotropy is 
$\approx 4\cdot 10^{-2}\cdot  \text {meV}$.
\end{thebibliography}
\end{document}


\setlength\abovedisplayshortskip{1pt}
\setlength\belowdisplayshortskip{1pt}
\setlength\abovedisplayskip{1pt}
\setlength\belowdisplayskip{1pt}
\title{Supplemental material\\ to\\ ''On the spontaneous magnetization of two-dimensional ferromagnets''}
\author{D. Pescia}
\affiliation{Laboratory for Solid State Physics, ETH Zurich, 8093 Zurich, Switzerland}
\email{pescia@solid.phys.ethz.ch}%
\author{A. Vindigni}
\affiliation{Laboratory for Solid State Physics, ETH Zurich, 8093 Zurich, Switzerland.}
\affiliation{Department of Evolutionary Biology and Environmental Studies, University of Zurich, 8057 Zurich, Switzerland.}
\maketitle
\section{Derivation of Eq.1: Magnetostatic energy of juxtaposed slabs with opposite perpendicular magnetization.}
\subsection{General considerations.}
\noindent According to \cite{Jak}, the magnetostatic self-energy of a continuous distribution of permanent magnetization $\vec M(\vec r)$ can be written as
\begin{eqnarray}
 E_M[\vec M(\vec r)]&=&-\frac{\mu_0}{2}\cdot \iiint dV \cdot \vec M(\vec r)\cdot \vec H(\vec r) -\frac{\mu_0}{2}\cdot \iiint dV \cdot \vec M^2(\mathbf r))
\end{eqnarray}
For the purpose of analyzing the situation of juxtaposed slabs with opposite perpendicular magnetization we use a sligthy different version for $E_M$. 
As $\vec B = \mu_0\cdot (\vec H + \vec M)$ 
and $\iiint dV\cdot \vec B(\vec r)\cdot \vec H(\vec r)= 0$
we can write  
\begin{equation}
E_M\! =\! -\frac{1}{2\mu_0}\iiint dV\cdot  \vec B(\vec r)\cdot \vec B(\vec r)
\end{equation} 
With $\vec B = \vec \nabla\times \vec A$ and the partial integration 
\begin{equation}\iiint_V \vec \nabla\times \vec A \cdot \left(\nabla \times \vec A\right)\,dV=-\int_{\partial V}\left(\nabla \times \vec A \right) \times \vec {A}\cdot d\mathbf{S} + \iiint_V \left(\nabla \times \nabla\times \vec A\right) \cdot \vec{A}\,dV
\end{equation}
we obtain, using $\vec \nabla\times \vec B  =\mu_0\cdot \vec \nabla\times \vec M$,
\begin{equation}
E_M[\vec M(\vec r)] = -\frac{1}{2\mu_0}\int dV\cdot \vec A\cdot \underbrace{\vec \nabla \times \vec \nabla \times \vec A}_{\mu_0\cdot \vec  \nabla\times \vec M}= -\frac{1}{2}\int dV\cdot \vec A\cdot \vec  \nabla\times \vec M
\end{equation}
(the surface terms are rendered vanishing by extending the surface to $\infty$, where the fields of a finite and bounded magnetization distribution are vanishing). We recall that $\vec A$ fulfills the Poisson  equation 
\begin{equation}
\triangle \vec A = -\mu_0\cdot \vec \nabla\times \vec M
\end{equation}
with solution 
\begin{equation}
\vec A(\vec r) = \frac{\mu_0}{4\pi}\cdot \int dV' \frac{\vec \nabla\times \vec M}{\vert \vec r -\vec r'\vert}
\end{equation}
Inserting this result in Eq.4 we obtain the sought-for representation of the total magnetostatic energy: 
\begin{equation}
E_M[\vec M(\vec r)]\!=\!-\frac{\mu_0}{8\pi}\iiint dV_1 \iiint dV_2 \frac{\vec \nabla_1\times \vec M(\vec r_1)\cdot \vec \nabla_2\times \vec M (\vec r_2)}{\vert \vec r_1 -\vec r_2\vert}
\end{equation} 
Eq.7 is the interaction energy of effective Amperian currents with current density vector 
\begin{equation}
\vec J_M \doteq \vec \nabla\times \vec M
\end{equation}
\subsection{Application to perpendicular magnetization in a slab.}
\noindent We compute explicitly the leading logarithmic contribution. Suppose we have two slabs with length $L$ along the $y$-direction, infinitely extended along the the $x$-direction and with thickness $d\!<\!<\!L$ along the $z$-direction. The slabs meet at $x=0$. Along this line, the perpendicular magnetization changes from $(0,0,M_0)$ to $(0,0,-M_0)$. This magnetization jump produces an effective current density: 
\begin{equation}
\vec \nabla \times \vec M = (0,\frac{2M_0}{w},0)
\end{equation}
localized at $x=0$. This expression entails the fact that, for physical reasons (see Section III), the line $x=0$ at which the domains meet is assigned a finite width $w\!<\!<\!L$. 
The self-energy of the effective current flowing along the domain wall amounts to 
\begin{equation}
-\frac{\mu_0}{8\pi}\cdot \frac{(2\cdot M_0)^2}{w^2}\iiint dV_1\iiint dV_2
\frac{1}{\sqrt{(x_1-x_2)^2+(y_1-y_2)^2+(z_1-z_2)^2}}
\end{equation}
The integral along $z$ extends from $0$ to $d$, the integral along $y$ from $0$ to
$L$ and the integral along $x$ from $0$ to $w$. 
We use the variables $\vec r'_1=\frac{\vec r_1}{L}$, $\vec r'_2=\frac{\vec r_2}{L}$
and transform the integral to 
\begin{equation}-\frac{\mu_0}{8\pi}\cdot \frac{(2\cdot M_0)^2}{w^2}\cdot L^5\cdot \iiint dV'_1\iiint dV'_2
\frac{1}{\sqrt{(x'_1-x'_2)^2+(y'_1-y'_2)^2+(z'_1-z'_2)^2}}
\end{equation}
The integral along $z'$ extends from $0$ to $\frac{d}{L}$, the integral along $y'$ from $0$ to
$1$ and the integral along $x'$ from $0$ to $\frac{w}{L}\doteq \omega$.
We solve this integral in the limit $\frac{d}{L}\!<<\!1$. In this limit, we set $z_1'=z_2' =0$ in the integrand and perform the $z'$-integral to obtain
\begin{equation}-\frac{\mu_0}{8\pi}\cdot \frac{(2\cdot M_0)^2}{w^2}\cdot\! L^5\!\cdot\! \frac{d^2}{L^2}\!\cdot\!\int_0^{\omega}\!dx'_1\!\int_0^{\omega}\!dx'_2\!\int_0^1\! dy'_1\!\int_0^1\!dy'_2\!\frac{1}{\sqrt{(x'_1-x'_2)^2+(y'_1-y'_2)^2}}
\end{equation}
The remaining integrals are elementary 
ones and the result of the exact integration is 
\begin{eqnarray}
-\frac{\mu_0}{8\pi}\cdot \frac{(2\cdot M_0)^2}{w^2}\cdot L^3\!\cdot\!d^2\!\cdot\! 2\cdot\!
\left\{\!\frac{\omega^2}{2}
\ln\left( \frac{\sqrt{1+\omega^2}+1}{\sqrt{1+\omega^2}-1}\right)
   \!+\! \omega\, \ln\left( \sqrt{1+\omega^2}+ \omega\right)
\!-\!\frac{1}{3}\left[ 
\left(1+\omega^2 \right)^{\frac{3}{2}}\!-\!\omega^3\!-1\!\right]\!
 \right\}
\end{eqnarray} 
We are interested in the situation where $w$ is also much smaller than $L$ while being larger than $d$. In this situation, Eq. 13 simplifies to 
 \begin{equation} 
\approx-\frac{\mu_0}{8\pi}\cdot \frac{(2\cdot M_0)^2}{w^2}\cdot L^3\!\cdot\!d^2\!\cdot\! 2\cdot\!
\left\{(\frac{w}{L})^2
\ln\left(\frac{1}{\frac{w}{L}}\right)  +(\frac{w}{L})^2  \left(\ln 2 +\frac{1}{2} \right) + 
\frac{(\frac{w}{L})^3}{3} +
\mathcal{O}\left((\frac{w}{L})^4\right)   \right\} 
\end{equation}
The leading term is the logarithmic one: 
\begin{equation}\approx-\frac{\mu_0}{4\pi}\cdot (2\cdot M_0\cdot)^2\cdot a^3\cdot \frac{L\cdot d^2}{a^3}\cdot \ln\frac{L}{w}
\end{equation}
We use the parameter
\begin{equation}
\Omega\doteq \frac{\mu_0}{2}\cdot M_0^2\cdot a^3
\end{equation}
to write this leading term as (see Eq.1 in the bulk of the paper)
\begin{equation}
-\frac{2}{\pi}\cdot (\Omega\cdot \frac{d}{a}) \cdot \frac{L\cdot d}{a^2}\cdot \ln\frac{L}{w}+ {\cal O}(\frac{d^2\cdot L}{a^3})
\end{equation}
For the sake of being close to practical examples, we compute $\Omega$ assuming  Fe atoms occupying a bcc lattice, i.e. 2 atoms in the unit cell, each carrying a magnetic moment of $2.2$ $\mu_B$ and  
\[\mu_0 = 4\pi\cdot 10^{-7} \cdot \frac{T\cdot m}{A}\quad \mu_B = 9.3\cdot 10^{-24}\cdot \frac{\text{Joule}}{T}\quad a= 2.83\cdot 10^{-10}\cdot m \quad 1\cdot \text{Joule} = 6.2\cdot 10^{18} \cdot eV\] 
We find
\begin{equation}
\Omega\approx 0.28\cdot \text{meV}
\end{equation}  
\section{The dipolar contribution to the Neel surface anisotropy: the term $\Omega\!\cdot\!d$.}
\subsection{General considerations.} For the purpose of dealing with this specific problem we insert 
\begin{equation}
\vec H = \frac{1}{4\pi}\cdot \iiint dV'\vec \nabla\frac{\vec \nabla' \cdot \vec M(\vec r')}{\vert \vec r-\vec r'\vert} 
\end{equation}
into
\begin{equation}
-\frac{\mu_0}{2}\cdot \iiint dV \cdot \vec M(\vec r)\cdot \vec H(\vec r)
\end{equation}   
We use the identity 
\begin{equation}
\vec M(\vec r')\cdot \vec \nabla \frac{1}{\mid \vec r -\vec r' \mid}= \vec \nabla^{'}\cdot \vec M(\vec r')\cdot \frac{1}{\mid \vec r -\vec r' \mid} - \vec \nabla^{'}\cdot (\vec M(\vec r')\cdot \frac{1}{\mid \vec r -\vec r' \mid})  
\end{equation} 
and suppose that $M(\vec r')$ is well behaved and localized. Then the integral over the last term can be transformed by Gauss law into a surface integral and the surface can be pushed to infinity, where there is no magnetization and the surface integral vanishes, leading to
\begin{equation}
E_M[\vec M(\vec r)]= +\frac{\mu_0}{8\pi}\iiint dV \iiint dV'\frac{\vec \nabla\cdot \vec M(\vec r)\cdot \vec \nabla^{'}\cdot \vec M(\vec r')}{\mid \vec r -\vec r' \mid}-\frac{\mu_0}{2}\iiint dV \cdot \vec M^2(\vec r) 
\end{equation}
\subsection{Application to the slab geometry.}
We consider a slab of size $L\times L $ in the $xy$-plane, finite thickness 
$d\!<\!<\!L$ along the $z$-direction. We need to compute the magnetostatic energy 
when the slab is filled with a uniform magnetization distribution $\vec M= 
(\sin\theta,0,\cos\theta)$ ($\theta$ being the angle with respect to the slab 
normal), subtracted by the magnetostatic energy at $\theta=\frac{\pi}{2}$. As 
we will let the slab extend to infinity, we proceed by setting the derivative of 
$\vec M(\vec r)$ along $x$ and $y$ to zero in Eq.22. The derivatives along $z$ can be redirected and we obtain, for the magnetic anisotropy energy $\Delta E_M\doteq E_M[\cos\theta]-E_M[\cos\frac{\pi}{2}]$ produced by the dipolar interaction 
{\small\begin{equation}
\Delta E_M= \cos^2\theta\cdot \frac{\mu_0}{8\pi}\cdot M_{0}^2\!\iiint\! d\vec \rho\!dz \iiint\!d\vec \rho'\!dz'
\left[\frac{\partial }{\partial z}\frac{\partial }{\partial z'}
\frac{1}{((\vec \rho-\vec \rho')^2 + (z-z')^2)^{1/2}}\right]
\end{equation}}
with $\vec \rho = (x,y)$.   
The integral over $z$ and $z'$ from $0$ to $d$ can be performed to obtain 
\begin{equation}
\Delta E_M=\cos^2\theta\cdot \frac{\mu_0}{4\pi}\cdot M_{0}^2\iint\!d\vec \rho\!\iint\!d\vec \rho'
\left[
\frac{1}{\sqrt{(\vec \rho-\vec \rho')^2}}-\frac{1}{\sqrt{(\vec \rho-\vec \rho')^2 + d^2}}\right]
\end{equation}
The integrations over the in-plane coordinates are elementary, as the integration limits in the plane can be considered to extend to $\infty$ and  $\vec \rho'$ can be set to zero:   
\begin{eqnarray}
\Delta E_M&=&\cos^2\theta\cdot \frac{\mu_0}{4\pi}\cdot M_{0}^2\underbrace{\int dx\cdot dy}_{L^2} \int dh\cdot dk \left[
\frac{1}{\sqrt{h^2+k^2}}-\frac{1}{\sqrt{h^2+k^2+d^2}}\right]\nonumber\\
&=& \cos^2\theta\cdot \frac{\mu_0}{4\pi}\cdot M_{0}^2\cdot L^2\cdot 2\pi\cdot \underbrace{\lim_{L\rightarrow \infty }\int_0^L dr(1-\frac{r}{\sqrt{r^2+d^2}})}_{d}\nonumber\\
&=& L^2\cdot \frac{\mu_0}{2}\cdot M_{0}^2\cdot d\cdot \cos^2\theta= 
(\frac{L}{a})^2\cdot (\Omega\cdot \frac{d}{a})\cdot \cos^2\theta
\end{eqnarray}
In this last equation one can read out that $\Omega\cdot \frac{d}{a}$ determines the coefficient of the magnetic anisotropy arising from the dipolar interaction.\\
One technical aspect: whithin a discrete model, ''one monolayer'' is defined by default \cite{GR}. Whithin the continuum slab model used here, one needs to define the value of $d$ corresponding to ''one monolayer''. One possible way of accomplishing this is comparing the coefficient $\Omega\cdot \frac{d}{a}$ in Eq.25 with the corresponding coefficient computed within the discrete model \cite{GR}. We find that setting $d\approx a$ for the ''one monolayer'' is indeed a suitable choice. 
\section{Derivation of Eq.2: The energy of a domain wall.}
For the exchange energy between a pair of neigboring spins at sites $\vec r_i$ and $\vec r_j$ we adopt the classical rendering which appears to be appropriate e.g. for metallic Fe, as explained in Ref.\cite{Small}: 
\begin{equation}
E_J = -J\cdot S^2\cdot \vec n(\vec r_i)\cdot \vec n(\vec r_j)
\end{equation}
$\vec n$ is a classical vector. $J$ is the exchange coupling constant given in Ref.\cite{Small} as $\approx 48$ meV. The use of $S^2$ instead of $S(S+1)$ is explained in Ref.\cite{Small}. For using this expression to computing the energy of a domain wall we need some approximations. 
Within a wall that evolves along the $x$-axis, $\vec r_i \doteq (x,0,0)$ and the $\vec r_j\doteq (x\pm a,0,0)$ and  
\begin{equation}
\vec n_i = (\sin\theta(x),0,\cos(\theta(x))\quad \vec n_j = (\sin\theta(x\pm a),0,\cos(\theta(x\pm a))
\end{equation}
The change of exchange energy produced by the misalignement of two consecutive spins amounts accordingly to  
\begin{equation}
E_J(\uparrow\downarrow)-E_J(\uparrow\uparrow)= 
-J\cdot S^2\cdot \cos\left(\theta(x)-\theta(x\pm a)\right)+ J\cdot S^2
\end{equation}
We expect that $\theta(x)-\theta(x\pm a)$ is small, so that we can replace the $\cos$ function with the lowest terms of its Taylor series and obtain
\begin{equation}
E_J(\uparrow\downarrow)-E_J(\uparrow\uparrow)\approx \frac{JS^2}{2}\left(\theta(x)-\theta(x\pm a)\right)^2
\end{equation}
We also use 
\begin{equation}
\left(\theta(x)-\theta(x\pm a)\right)^2\approx \frac{\partial \theta(x)}{\partial x}^2\cdot a^2
\end{equation}
This approximation allow to write the exchange component of the total elastic functional that is used to compute the equilibrium profile of $\theta(x)$. 
Together with the term originating from the Neel \cite{Neel} magnetic anisotroy, the functional $E_w[\theta(x)]$ describing the total energy of a wall writes  
{\small \begin{equation}
E_w[\theta(x)]=\frac{L\cdot d}{a^2}\cdot\left[\frac{J\cdot S^2}{2}\cdot a\cdot \int_{-\infty}^{\infty} \left(\frac{\partial \theta(x)}{\partial x}\right)^2 \cdot dx-(\lambda-\Omega\cdot \frac{d}{a})\frac{1}{a}\int dx\cdot \cos^2\theta(x)\right]
\end{equation}}
The corresponding Euler-Lagrange equation reads 
\begin{equation} 
J\cdot a \frac{d^2 \theta }{d x^2} 
-2\frac{\lambda-\Omega\frac{d}{a}}{a} \sin( \theta(x))\cos( \theta(x))=0
\end{equation}
The solution to the boundary conditions $\lim_{x\rightarrow -\infty} \theta(x) = \pi$ and $\lim_{x\rightarrow +\infty} \theta(x) = 0$
reads
\begin{equation} 
\cos(\theta(x)) = \tanh\left(\frac{x}{w}\right) 
\end{equation}
with the width $w$ of the wall 
\begin{equation}
w=\frac{a}{2}\sqrt{\frac{2\cdot J\cdot S^2}{\lambda-\Omega\frac{d}{a}}}
\end{equation}
This solution was proposed originally by Landau and Lifschitz in 1935 \cite{LL}. 
For the sake of finding the proper role of the dipolar interaction within the expression for the wall energy we have repeated this solution here. The total energy of the wall amounts to (see Eq.2 in the 
bulk of the paper)
\begin{equation}
\frac{L\cdot d}{a^2}\cdot 2\cdot \sqrt{\left(\lambda-\Omega\frac{d}{a}\right)\cdot 2\cdot J\cdot S^2}\doteq E_w
\end{equation}
\section{Stability of a stripe domain in an applied magnetic field.}
We have determined that, when the size of a magnetic element exceeds  the 
crossover length $L_c$, a phase with domains of opposite perpendicular 
magnetization can penetrate a magnetic element. We now determine the stability 
of this phase with respect to a magnetic field applied perpendicularly to the 
plane. For this purpose, we compute the magnetic field 
necessary to render metastable one stripe of reversed magnetization with width $\delta$. The change in total energy produced by the stripe with respect to an element of uniform magnetization $+M_0$ in an applied magnetic field $\vec B = (0,0,+B_0)$ amounts to 
\begin{equation}
\Delta E(B_0,\delta)= 2\cdot E_w\cdot \frac{L\cdot d}{a^2} + 2\cdot B_0\cdot M_0\cdot\delta\cdot L\cdot d-\frac{4}{\pi}\cdot (\Omega\cdot\frac{d}{a})\cdot \frac{L\cdot d}{a^2}\cdot \ln\frac{\delta}{w} + {\cal O}(\frac{L\cdot d}{a^2}\cdot \Omega\cdot\frac{d}{a})
\end{equation}
The $-\ln\frac{\delta}{w}$ in Eq. 36 comes about by subtracting $2\cdot \ln\frac{L}{w}$ (originating from the the self energy of the effective current densities flowing along the domain walls) from $2\cdot \ln \frac{L}{\delta}$ (originating from the reciprocal interaction energy between the effective current densities flowing along the domain walls \cite{foot}). 
Minimizing $\Delta E(B_0,\delta)$ with respect to $\delta$ produces the equilibrium stripe width
\begin{equation}
\delta^*(B_0)= a\cdot \frac{\frac{2}{\pi}\cdot \Omega\cdot \frac{d}{a}}{B_0\cdot M_0\cdot a^3} = d \cdot \frac{\mu_0}{\pi} \cdot \frac{M_0}{B_0} 
\end{equation}
On p.20-21 of Ref.\cite{Niculin}, a numerical study of $\Delta E(B_0,\delta)$ as a function of $\delta$ for various fields $B_0$ is reported. It is shown that, below a certain threshold magnetic field $B_t$ (and, in particular, at $B_0=0$), this minimum renders $\Delta E(B_0,\delta)$ negative: the energy of the element containing the stripe is lower than the energy of the element with uniform magnetization. Above the threshold field, instead, the minimum 
provides a metastable state, as $\Delta E(B_0,\delta^*)>0$ \cite{Niculin}. We now proceed to find $B_t$. To find $B_t$ we insert the expression for $\delta^*(B_0)$ in $\Delta E(B_0,\delta)$, i.e. we build  the function $\Delta E(B_0,\delta^*(B_0))$:
\begin{equation}
\Delta E(B_0,\delta^*(B_0))= \frac{L\cdot d}{a^2}\cdot\left(2\cdot E_w -\frac{4}{\pi}\cdot (\Omega\cdot\frac{d}{a})\cdot \ln\left(\frac{\mu_0\cdot M_0}{\pi\cdot B_0}\cdot \frac{d}{w}\right) + {\cal O}(\Omega\cdot \frac{d}{a})\right) 
\end{equation}
$\Delta E(B_0,\delta^*(B_0))$ expresses the energy change at the minimum $\delta^*$ as a function of $B_0$. Setting $\Delta E(B_0,\delta^*(B_0))$ to zero provides us with an equation in the variable $B_0$ for the sought-for field $B_t$. The solution of this equation  writes 
\begin{equation}
B_t\propto \mu_0\cdot M_0\cdot \frac{d}{L_c}
\end{equation}
This is the transition field referred to in the bulk of the paper.
A similar dependence of $B_t$ on $M_0$ and on the ratio $L_c/d$ was obtained in a situation of stripe order in 2D\cite{Niculin,NiculinPRB}. Notice that, even if the stripe is metastable, there is an energy barrier for it to be eliminated from 
the element.  A further issue is that a bubble phase might 
intervene between the stripe phase and the phase of uniform magnetization. 
For the purpose of this paper, however, these issues are less relevant and we refer the reader to Refs.\onlinecite{Niculin,NiculinPRB} for an extended discussion.  

\section{Magnetostatic energy of juxtaposed  domains with opposite in-plane  magnetization.}
For the in-plane configuration we compute the magnetostatic energy using 
\begin{equation}
E_M[\vec M(\vec r)] = +\frac{\mu_0}{8\pi}\int dV \int dV'\frac{\vec \nabla\cdot \vec M(\vec r)\cdot \vec \nabla^{'}\cdot \vec M(\vec r')}{\mid \vec r -\vec r' \mid}
\end{equation}
(the summand on the right hand side of the exact Eq.22 cancels out when energies of two different configurations with equal $\iiint dV \vec M^2(\vec x)$ are subtracted). 
We consider the two magnetization distributions ''\textit{uu}'' (a in the top view of the slab in Fig.1SM) and ''ud'' (b in Fig.1SM) (for a magnetization pointing along $+x$ and $-x$ we use the symbols ''u'' and ''d'', respectively). In \textit{uu} the in-plane magnetization $\vec M = (0,M_0,0)$ is uniformly distributed. It produces a $\vec \nabla\cdot \vec M$ in the vicinity of the segments terminating the slabs: effective negative charges accumulate along the segments A and B   and effective positive charges along the segments C and D, see Fig.1SM. In \textit{ud}, half of the slab is filled with $\vec M = (0,+M_0,0)$ and the remaining half with the opposite magnetization $(0,-M_0,0)$. The effective charges at the terminating segments are accordingly modified: A and D are negatively charged, B and C positively. In a situation where the domain boundary is exactly  parallel to the magnetization vector, the self-energy of the charge distributions cancel out when the magnetostatic energy of the configuration \textit{uu} is subtracted from the energy of the configuration \textit{ud}. The remaining terms amount to the Coulomb interaction (symbolically)  $-4\cdot A^{(+)}B^{(+)} + 4\cdot A^{(+)}D^{(+)}$. 
The leading term is the Coulomb energy between the charges along the segments $A$
and $B$, and it is negative, i.e. the formation of domains with parallel opposite magnetization lowers the magnetostatic energy of the slab. In the evaluation of this energy, the integration along $z$ produces a $d^2$ by 
simultaneously setting $z=z'=0$ in the integrand. The integration along $y$ 
produces $\Delta^2$ by simultaneously setting $y=y'=0$ in the integrand.
 $\Delta$  is the length of that border region at $A,B,C,D$ across which the magnetization decays to $0$. The integration over $x$ and $x'$ must be performed explicitly. We insert a wall of finite thickness $w$ between the two domains. At the end of the calculation we will let $w$ go to zero. This will show that this energy contribution is also finite. $E(ud)-E(uu)$ writes, approximately    
\begin{eqnarray}
E(ud)-E(uu)&\approx &(-4)\cdot \frac{\mu_0}{8\pi}M_0^2\cdot d^2\cdot \int_{-L}^{-w} dx'\int_w^L dx \frac{1}{\vert x-x'\vert}\nonumber\\
&=& (-4)\cdot \frac{\mu_0}{8\pi}M_0^2\cdot d^2\cdot L\int_{-1}^{-\frac{w}{L}}dx'\int_{\frac{w}{L}}^1 dx \frac{1}{\vert x-x'\vert}\nonumber\\
&=&(-4)\cdot \frac{\mu_0}{8\pi}M_0^2\cdot d^2\cdot L\int_{-1}^{-\frac{w}{L}}dx'\ln\frac{1-x'}{\frac{w}{L}-x'}\nonumber\\
&\underbrace{\approx}_{w\rightarrow \infty} & -4\cdot 2\cdot \ln 2\cdot \frac{\mu_0}{8\pi}M_0^2\cdot d^2\cdot L= -\frac{2\ln 2}{\pi}\cdot \Omega\cdot \frac{d^2\cdot L}{a^3}
\end{eqnarray}
This result is used to write Eq.4 in the bulk of the paper.

\clearpage
\begin{figure}
\begin{center}
\includegraphics*[width=0.4\textwidth]{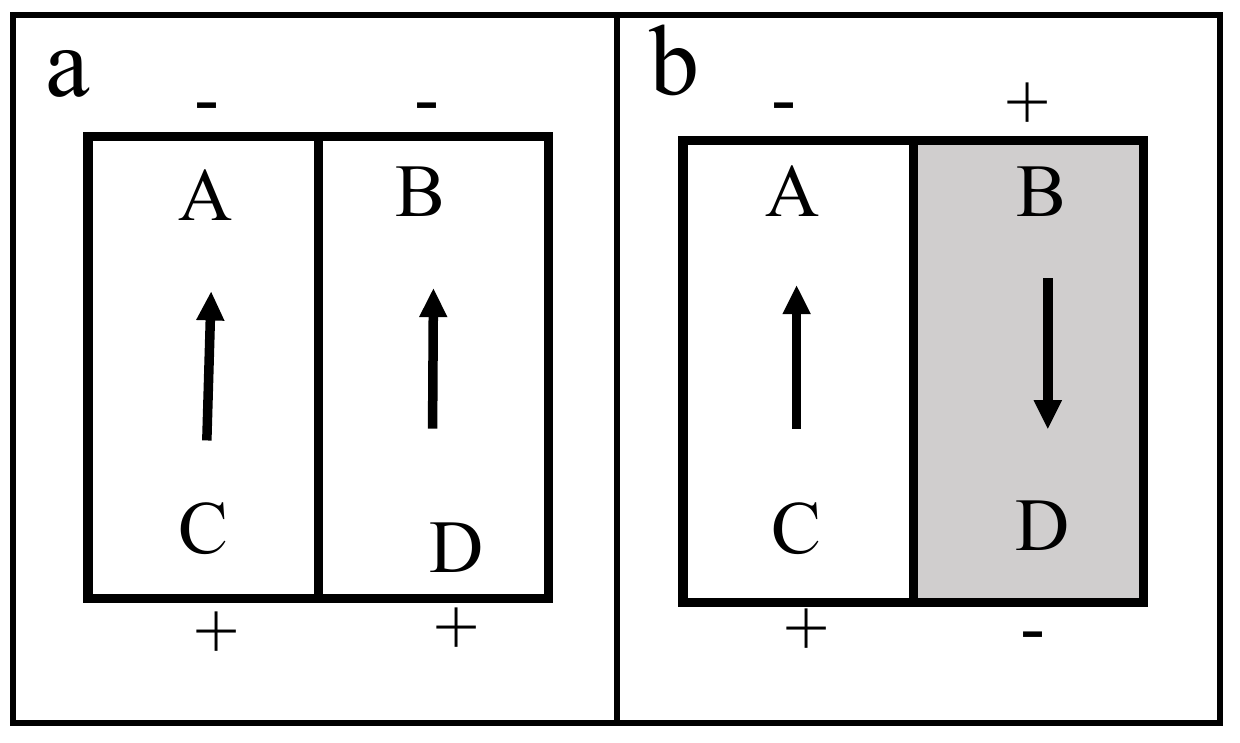}
\caption{a: Top view of the slab in the state of uniform in-plane magnetization (represented in white). Black arrows represent the magnetization vector. The sign of the effective charges appearing along the segments A,B,C,D is given. b. Top view of the slab with two domains of opposite in plane magnetization (white and gray), magnetization vectors represented by black arrows. The sign of the effective charges appearing along the segments A,B,C,D is given.  
}
\end{center}
\end{figure}